\documentclass{iau}
\usepackage{graphicx}
\usepackage{epstopdf}
\usepackage{natbib}

\title[Black Hole masses of LSB galaxies] 
{Investigating the AGN activity and black hole masses in Low Surface brightness galaxies}

\author[Subramanian et. al.]   
{Smitha Subramanian$^1$., Ramya.S.$^2$, Mousumi Das$^1$., Koshy George$^1$., T. Sivarani$^1$ 
 \and T.P.Prabhu$^1$}

\affiliation{$^1$ Indian Institute of Astrophysics, 2nd Block, Koramangala, Bangalore - 560034, India
 \\ email: {\tt smitha@iiap.res.in,mosumi@iiap.res.in,koshy@iiap.res.in,sivarani@iiap.res.in,tpp@iiap.res.in} \\[\affilskip]
$^2$Shanghai Astronomical Observatory, Shanghai, China \\email: {\tt sramyaraman@gmail.com}}

\pubyear{2014}
\volume{xxx}  
\pagerange{119--126}
\jname{Star Clusters and Black Holes in galaxies across cosmic time}
\editors{A.C. Editor, B.D. Editor \& C.E. Editor, eds.}
\begin{document}

\maketitle

\begin{abstract}
We present an analysis of the optical nuclear spectra from the active galactic nuclei (AGN) in a sample of giant low surface brightness 
(GLSB) galaxies. GLSB galaxies are extreme late type spirals that are large, isolated 
and poorly evolved compared to regular spiral galaxies. 
Earlier studies have indicated that their nuclei have relatively low mass black holes.
Using data from the Sloan Digital Sky Survey (SDSS), we selected a sample of 30 GLSB galaxies that showed 
broad H$\alpha$ emission lines in their AGN spectra. In some galaxies such as UGC 6284, 
the broad component of H$\alpha$ is more related to outflows rather than the black hole. One 
galaxy (UGC 6614) showed two broad components in H$\alpha$, one associated with the black hole and the other associated 
with an outflow event. We derived the nuclear black hole (BH) masses of 29 galaxies from their broad H$\alpha$ parameters. We find that the nuclear BH 
masses lie in the range $10^{5}-10^{7}~M_{\odot}$. The bulge stellar velocity dispersion $\sigma_{e}$ was determined from the underlying 
stellar spectra. We compared our results with the existing BH mass - velocity dispersion ($M_{BH}-\sigma_{e}$) correlations and found that 
the majority of our sample lie in the low BH mass regime and below the $M_{BH}-\sigma_{e}$ correlation. 
The effects of galaxy orientation in the measurement of $\sigma_e$ and the increase of $\sigma_e$ due to the effects of bar 
 are probable reasons for the observed offset for some galaxies, but in many galaxies the offset is real. 
A possible explanation for the $M_{BH}-\sigma_{e}$ offset could be lack of mergers and 
accretion events in the history of these galaxies which leads to a lack of BH-bulge co-evolution. 
\end{abstract}
\section{Introduction}

Low surface brightness (LSB) galaxies are late type spiral galaxies (Sc or Sd) that have
a central disk surface brightness of $\mu_{B(0)}~\ge~$22 to 23 mag arcsec$^{-2}$
(\citealt{ib97}; \citealt{ibs01}). 
Disk LSBs are more rare and usually found in isolated environments \citep{b93} and vary over
a range of sizes.
But the really large LSB galaxies, which are
referred to as giant LSB (GLSB) galaxies, are generally found to lie close to the edges of voids
\citep{r09}. Nuclear activity is not common in LSB galaxies. This is in marked contrast to high surface 
brightness disk galaxies where the percentage having active galactic nuclei (AGN) can be as high as 
50$\%$, depending on the mean luminosity of the sample. The most probable explanation for this low fraction 
is that 
LSB disks generally lack two structural features that facilitate gas flows and the formation of a 
compact object in the nucleus - bars and strong spiral arms \citep{bim97}.
 
However, a significant fraction of bulge dominated GLSB galaxies show AGN activity \citep{s95}. 
AGN activity in bulge dominated GLSBs is not suprising as 
studies indicate that the growth of
nuclear black holes (BH) in galaxies is intimately linked to the growth of their bulges (\citealt{sr98}; 
\citealt{h04}). The strong correlation of black hole mass (M$_{BH}$) with bulge stellar velocity dispersion
 in galaxies (M$_{BH}$ - $\sigma_e$) is due to this supermassive black hole (SMBH) - bugle
co-evolution (\citealt{g09} and \citealt{mm13}). But in LSB galaxies, their 
bulge velocity dispersion 
and disc rotation speeds suggest that they lie 
below the M$_{BH}$ - $\sigma_e$ correlation for bright galaxies \citep{p05}. 
\cite{m09} estimated the black hloe masses of three GLSB 
galaxies and found them to lie close 
to the M$_{BH}$ - $\sigma_e$ correlation and the BH masses are found to be of the order of $\sim$ 10$^7$ 
M$_{\odot}$. 
Later estimates from the study of another three GLSBs  
by \cite{r11} showed that their sample galaxies lie offset from the M$_{BH}$ - $\sigma_e$ correlation. 
They suggested 
that the bulges of their sample might be well evolved, but the BH masses (3-9 x 10$^5$ M$_{\odot}$) 
are lower than those found 
in bright galaxies. \cite{m12} found that the bulges of LSB glaxies have similar properties 
as that of the bulges of normal galaxies. It is not clearly understood where the LSB galaxies in general 
lie on the M$_{BH}$-$\sigma_e$ correlation and also what are the possible reasons for the observed offset 
in a few sample galaxies. 
Low mass black holes in isolated LSB galaxies are are also very interesting candidates for the study 
of seed black holes in galaxies \citep{v10}.

\vspace{-0.3cm}
\section{Data and Analysis}
As an initial sample we selected 558 LSB galaxies from the literature (\citealt{s95}, \citealt{i96}, 
\citealt{s92} and \citealt{s88}) for which the Sloan Digital Sky Survey (SDSS) DR10 nuclear spectra are available. The emission line 
fluxes of these 558 galaxies are available for public in the SDSS DR10 database. 
The emission line flxes are used to construct the [N{\sc ii}]$_{6583}$/H${\alpha}$ vs [O{\sc iii}]$_{5007}$/H${\beta}$ 
Baldwin-Phillips-Terlevich (BPT) \citep{bpt81} diagram. Out of 558 galaxies, 160 
galaxies ($\sim$29\%, with almost equal contribution from the 
composite and purely AGN candidates) are classified as AGN and the remaining 398 ($\sim$71\%) galaxies 
as purely star forming systems. 
From the sample of 160 galaxies, we selected only those galaxies which have median S/N $>$ 15 for further analysis. 
Thus, finally we used SDSS DR10 spectra of 115 LSB galaxies to identify sources 
with broad balmer lines and hence estimate their black hole masses. 

We used the pPXF (Penalized Pixel-Fitting stellar kinematics 
extraction) code by \cite{ce04} to obtain the best fit model for the underlying stellar 
population. pPXF also provides the stellar velocity dispersion, $\sigma$. 
The H$\alpha$ + [N{\sc ii}] doublet region of all the 115 galaxies are first fitted with three Gaussians, 
representing the three narrow lines of H$\alpha$ and [N{\sc ii}] doublet. Then this region is again fitted including 
a fourth Gaussian component for the broad H$\alpha$ line. 
The final 
sample with broad H$\alpha$ component is selected based on the two criteria, that the the inclusion of the broad 
component should improve the reduced $\chi^2$ value by at least 5\% and also the broad H$\alpha$ peak flux should 
be at least three times larger than the residue of the fit. Thus based on the these criteria, 52 LSB galaxies 
were selected as galaxies with broad H$\alpha$ component. After the careful examination of the fits of each of these 
galaxies, 28 of them showed the presence of broad H$\alpha$ at significant level and 2 galaxies showed an extra 
broad components, which are significantly blue/red shifted from the H$\alpha$ central wavelength and 
are hence more likely to be associated  with an outflow rather than with the active black hole.  
The 28 galaxies, with genuine broad H$\alpha$ component, are of our interest as their black hole masses 
can be estimated and hence check their M-$\sigma_e$ correlation. UGC 6614, which 
showed signatures of an outflow, has a clear broad H$\alpha$ component along with the blue shifted 
outflow component.  The other emission lines present in the spectra of these 30 galaxies (28 with broad H$\alpha$ component and 2 with 
outflow signatures) were also 
analysed and fitted with Gaussian profiles. 
IDL programs are used to fit the emission line profiles with Gaussian functions and the fluxes of all the emission lines are estimated.  
\vspace{-0.3cm}
\section{Black Hole masses and M-$\sigma_e$ correlation}
The virial black hole masses of 29 broad line AGN candidates are calculated using the equation given in \cite{r13}, using the luminosity and FWHM of broad H$\alpha$. The equation is given below. Here, the 
scale factor that depends on the broad line region geometry, $\epsilon$ is assumed to be 1.  \\
$log {M_{BH} \over M_{\odot}} =  6.57 + 0.47 log {L_{H\alpha} \over 10^{42} ergs s^{-1}} + 2.06 
log {FWHM_{H\alpha} \over 10^{3} km s^{-1}}$\\
The masses estimated are in the range 5.5 x 10$^5$ M$_{\odot}$ - 3.5 x 10$^7$ M$_{\odot}$. The median mass is found 
to be 5.6 x 10$^6$ M$_{\odot}$. The median mass suggests that the LSB galaxies have black hole masses which 
are slightly higher than intermediate mass black holes. 

\begin{figure}[h]
 \begin{center}
  \includegraphics[width=3.4in]{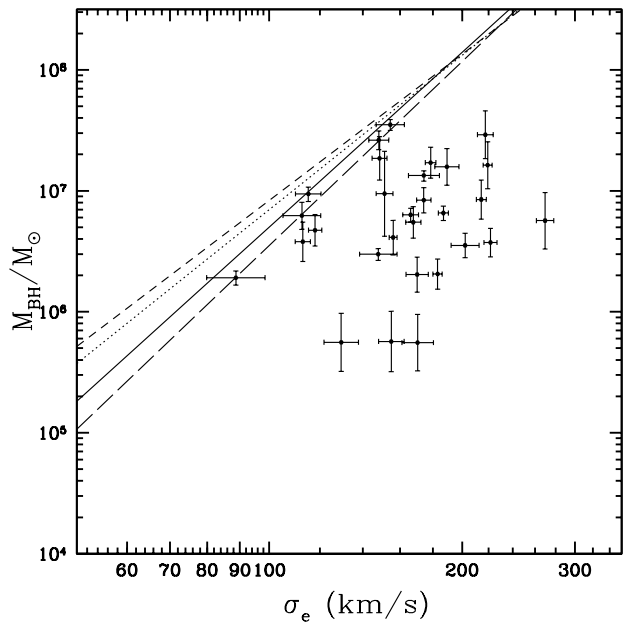} 
 \caption{The M-$\sigma_e$ plot with broad line AGN candidates. The linear regression lines given by 
\cite{t02}, \cite{fm02}, \cite{g09} and \cite{mm13} relation for late type 
galaxies (dashed, solid, dotted and long-dashed lines, respectively) for M$_{BH}$ against $\sigma_e$ are also 
shown.}
   \label{fig1}
 \end{center}
 \end{figure}

The stellar velocity dispersion values obtained from pPXF were transformed to the equivalent velocity
dispersion $\sigma_e$
at a radius of $r_e$/8 in the galaxy bulge, where $r_e$ is the bulge exponential scale length, using the
transformation equation given by \cite{j95}. The BH mass estimates for the 29 galaxies in our sample
plotted against the equivalent velocity dispersion values, $\sigma_e$ are shown in 
the M -$\sigma_e$ plot in Fig. 1. Also plotted
in the figure are the linear regression lines given by \cite{t02}, \cite{fm02}, \cite{g09} and 
\cite{mm13} for 
late type galaxies (dashed, solid, dotted and long-dashed lines, respectively) for M$_{BH}$ against $\sigma_e$. Most of 
our sample lie below the M-$\sigma_e$ correlation. The probable reasons for the observed offset 
can be due the the effects of galaxy orientation in the measurement of $\sigma$ and/or the effects of bars in 
the increase of $\sigma$ in case of barred galaxies. When analysed, we found that even if we incorporate the upper 
limits of these effects, they are not sufficent to explain the observed offset in more than 40\% of 
our sample. 
\vspace{-0.4cm}
\section{Discussion}
Seed BHs that are formed in the early universe grow by mass accretion
to become the massive BHs that we observe in our local universe; the accretion is driven by galaxy mergers 
and interactions. LSB galaxies are usually isolated and lie close to the edge of voids. Simulations suggest that 
lighter seed BHs grow through slow accreation leading relatively low mass BHs and lie 
below the M$_{BH}$-$\sigma_e$ correlation. These low mass 
and relatively pristine black holes can reveal important clues to the initial black hole mass function 
and help us to constrain the early evolution of black holes in galaxies. The black hole masses in our sample lie 
in the range 10$^5$ - 10$^7$ M$_{\odot}$ and most of them lie below the M$_{BH}$-$\sigma_e$ correlation. 
Also, their Eddington ratios are not high, hence their black holes 
are not accreting 
at a high rate and they fall in the low luminosity AGN (LLAGN) class.
Thus, the BHs in GLSB
galaxies represent one of the best candidates for pristine black holes and a good place to study seed black 
holes in our local universe. 
\vspace{-0.3cm}
\acknowledgements
The authors would like to thank the SDSS team for making the data publicly available. SS acknowledges the financial support from IAU for 
the participation in the symposium.  
%
%

\end{document}